\documentclass[11pt]{article} 

\usepackage[utf8]{inputenc} 

\usepackage{geometry} 
\usepackage{graphicx} 

\usepackage{booktabs} 
\usepackage{array} 
\usepackage{paralist} 
\usepackage{verbatim} 
\usepackage{subfig} 

\usepackage{fancyhdr} 
\usepackage{sectsty}
\allsectionsfont{\sffamily\mdseries\upshape} 

\usepackage[nottoc,notlof,notlot]{tocbibind} 
\usepackage[titles,subfigure]{tocloft} 

\usepackage{csquotes}
\usepackage{float}
\newfloat{exhibit}

\title{Power-law Portfolios}
\author{Jan Rosenzweig}
\date{}

\begin{document}

\maketitle

\abstract{Portfolio optimization methods suffer from a catalogue of known problems, mainly due to the facts that pair correlations of asset returns are unstable, and that extremal risk measures such as maximum drawdown are difficult to predict due to the non-Gaussianity of portfolio returns. \\
In order to look at optimal portfolios for arbitrary risk penalty functions, we construct portfolio shapes where the penalty is proportional to a moment of the returns of arbitrary order $p>2$. \\
The resulting component weight in the portfolio scales sub-linearly with its return, with the power-law $w \propto \mu^{1/(p-1)}$. This leads to significantly improved diversification when compared to Kelly portfolios, due to the dilution of the winner-takes-all effect.\\
In the limit of penalty order $p\rightarrow\infty$, we recover the simple trading heuristic whereby assets are allocated a fixed positive weight when their return exceeds the hurdle rate, and zero otherwise. Infinite order power-law portfolios thus fall into the class of perfectly diversified portfolios.\\}


{\bf Key takeaways:}
\begin{itemize}
\item power-law portfolios  address a number of know problems of covariance-based portfolios
\item weights of components scale sub-linearly with their performance, reducing portfolio concentration and the winner-takes-all problem
\item portfolio diversification improves with the increasing order of penalty, ultimately leading to a perfectly diversified portfolio as the penalty order tends to infinity.
\end{itemize}

\section{Introduction}

Textbook portfolio construction starts with the assumption that all assets are Gaussian, with  perfectly known returns, variances and correlations. It then proceeds to apply the Kelly criterion to construct the Markowitz mean-variance portfolio through it mass-adoption variant, the Capital Asset Pricing Model. The resulting portfolio is strongly dependent on the returns, variances and covariances. If the components were indeed Gaussian and the parameters were determined to a sufficient degree of accuracy, it would be maximising the Sharpe ratio for the given asset universe \cite{lintner, litterman, mark1, mark2, sharpe, tobin}. 

In the real world, the components are not Gaussian, and their returns, variances and correlations are, at best, rough estimates. This leads to a number of well-known problems in portfolio construction, and a series of various partial solutions to those problems \cite{avellaneda, shkolnik, tanzohren}. 

At the core of our approach is the simple notion that, while Gaussian random variables are completely determined by their return and covariance, non-Gaussian random variables carry non-trivial information at all moments. 

Similarly to \cite{rosenzweig}, we therefore proceed to look into portfolio construction when the risk penalty is attributed some arbitrary high order moment, rather than just the covariance. 

A number of studies have looked at various parametric non-Gaussian random variables and various risk-based penalty functions \cite{ahmadi, follmer, cajas, giller}.

We, on the other hand, make no specific assumptions about either the form of the underlying random variables, nor the specific risk measure involved. Rather, we seek to understand how different moments of the joint distribution influence the resulting portfolio construction. 

We recover a simple scaling law relating the component weight to its return. We also look at what happens when the order of the penalty movement goes to inifinity, and in this limit we recover the well nown trading heuristic  \cite{giller}
 \begin{displayquote}
\emph{Don’t trade if the signal is too small. If it is large enough, buy a fixed size.}
\end{displayquote}

The paper is set out as follows. Second section addresses the power-law portfolio construction in general. Third section dealis with the infinite order limit.   Fourth section looks at what happens if the underlying variables were actually Gaussian. Fifth section optimizes a portfolio of S\&P 500 stocks over the period of 12 years using power-law portfolios for values of $p$ varying over orders of magnitude. The final section discusses the results.

\section{Power-law weights}

Let $S^{(1)}_{t}..S^{(N)}_{t}$ denote prices of $N$ assets  at time $t$, forming a vector of asset prices ${\bf S}_{t}$, and $d{\bf S}_{t}$ is the vector of its increments.

We denote by
$${\bf m}_{t} = E\left(d{\bf S}\right)$$
the mean return of the joint distribution of $d{\bf S}_{t}$, and by ${\bf r}_{t}$ the vector of funding rates.

The portfolio optimization problem that we are interested in is the selection of a normalized weights vector ${\bf w}$ such that
\begin{equation}
 {\bf w}.({\bf m}_{t} - {\bf r}_{t}) - \lambda\  E\left| {\bf w} . (d{\bf S}_{t}  - {\bf m}_{t}) \right|^{p} \rightarrow \max 
\label{optimization}
\end{equation}
for  some choice of $\lambda, p>0$.
In other words, we are looking for portfolio weights that maximize the return of the portfolio, while penalizing for its absolute moment of order $p$. 

There is a closely  a related problem, 
\begin{equation}
 {\bf w}.({\bf m}_{t} - {\bf r}_{t}) - \lambda\ \left|  E\left( {\bf w} . (d{\bf S}_{t}  - {\bf m}_{t}) \right)^{p} \right| \rightarrow \max 
\label{optimization1}
\end{equation}
where the penalty term is the absolute value of the signed central moment, as opposed to the absolute central moment.

We are primarly motivated by the cases where $p$ is an even integer, so that both formulations are exactly the same. We can, however, expand the results to other positive values of $p$.

Following  \cite{rosenzweig}, we solve it using the Independent Component Analysis (ICA) \cite{ hyva}. Given the independent components $IC^{(1)},... IC^{(N)}$ with respective means $\mu^{(i)}$, funding rates $r^{(i)}$,  $p$th  absolute central moments $M_{p}^{(i)}$ and  $p$th  signed central moments $m_{p}^{(i)}$, we look for the solution in the form
\begin{equation}
 \Pi \propto \sum_{i} w^{(i)} IC^{(i)} .   \label{ICAportfolio}
\end{equation}

Substituting (\ref{ICAportfolio}) into (\ref{optimization}), we can always find a local maximum, which is  under reasonable conditions also the global maximum of (\ref{optimization}) (see Appendix); it has the asymptotic expansion  
\begin{equation}
w^{(i)} \propto \left\{ 
\begin{array}{ll}
0 & \mu^{(i)} - r^{(i)} \le 0 ; \\
 \left( \frac{\mu^{(i)} - r^{(i)}}{M_{p}^{(i)}} \right)^{1/(p-1)} + C_{0}^{(i)} +  C_{1}^{(i)}\left( \mu^{(i)} - r^{(i)} \right)^{-1/(p-1)}  +... & \texttt{otherwise.}
\end{array} \right.
   \label{ICAweightsnohurdle}
\end{equation}
for the formulation (\ref{optimization}), and 
\begin{equation}
w^{(i)} \propto \left\{ 
\begin{array}{ll}
0 & \mu^{(i)} - r^{(i)} \le 0 ; \\
 \left( \frac{\mu^{(i)} - r^{(i)}}{|m_{p}^{(i)}|} \right)^{1/(p-1)} + C_{0}^{(i)} +  C_{1}^{(i)}\left( \mu^{(i)} - r^{(i)} \right)^{-1/(p-1)}  +... & \texttt{otherwise.}
\end{array} \right.
   \label{ICAweightsnohurdle1}
\end{equation}
for the formulation (\ref{optimization1}).

In particular, the leading order term in (\ref{ICAweightsnohurdle}) is separable, i.e. $w^{(i)}$ depends only on the properties of $IC^{(i)}$ at leading order, and not on any of the  $IC^{(j)}$ for $j\ne i$. Separability does not extend to  higher order coefficients $C_{m}^{(i)}$, which generally do depend on  $IC^{(j)}$ for $j\ne i$.

In practice, the leading order term is sufficient in  most cases, leading to the approximate component weight formulas

\begin{equation}
w^{(i)} \propto \left( \frac{\mu^{(i)}-r^{(i)}}{M_{p}^{(i)}} \right)^{1/(p-1)}    \label{ICAweights}
\end{equation}
\begin{equation}
w^{(i)} \propto \left( \frac{\mu^{(i)}-r^{(i)}}{|m_{p}^{(i)}|} \right)^{1/(p-1)}    \label{ICAweights1}
\end{equation}
for problems (\ref{optimization}) and (\ref{optimization1}), respectively.

The  case of $p=2$ is well known; equation (\ref{ICAweights}) then becomes  the well known Kelly criterion for maximising the Sharpe ratio of the portfolio \cite{lintner, litterman, mark1, mark2, sharpe}. Each component is normalised by its volatility, and then weighted in proportion to its Sharpe ratio.

The case of $p=4$ has also been studied before; $M_{4}$ is the kurtosis of the component, and the exponent is $1/3$;  we therefore recover the Fat-tailed ratio of Rosenzweig  \cite{rosenzweig}, which maximizes the ratio of portfolio return to its kurtosis.

Intuitively, as discussed above, we are motivated by the cases where $p$ is an even integer, so that the absolute central moment and the signed central moment are exactly the same. The weights formula (\ref{ICAweights}), however, obviously works for any non-integer $p \ne 1$, as long as the moment in the penalty function is the absolute central  moment. If $p\le 1$, the penalty grows slower than the gain, and (\ref{optimization}) is optimized by setting all weights to zero.

With that in mind, are restricting ourselves to $p\ge 2$, while allowing $p$ to be non-integer. It is, however, important to note that, if $p$ is not an even integer, further technical conditions are needed to ensure that the asymptotic expansion (\ref{ICAweightsnohurdle})  approximates the global maximum of (\ref{optimization}).

The most interesting feature of (\ref{ICAweights}) is the sub-linear scaling of the component weight with its performance. In Sharpe-maximizing Kelly portfolios, the weight is proportional to the performance. A component with twice the performace of another will generally receive twice its weight. 

Formula (\ref{ICAweights}) shows, however, that that is  a special case. For any choice of $p>2$, the scaling is  sub-linear, and the component with twice the performance will receive less than twice the weight. This, in turn, directly addresses the winner-takes-all problem inherent in Kelly portfolios, whereby the portfolio is dominated by a small number of highly performing components, which in turn negates the benefits of diversification.

Using a higher order penalty for portfolio construction as per (\ref{optimization}) directly addresses this issue, by effectively limiting the over-weighting of higly performing components.

\section{Infinite Order Penalty}

The most interesting result of the previous section concerns the sub-linear dependence of each component weight on its return. This becomes more prominent as the penalty order $p$ becomes larger.

\begin{figure}[p]
\centering
\includegraphics[width=10cm]{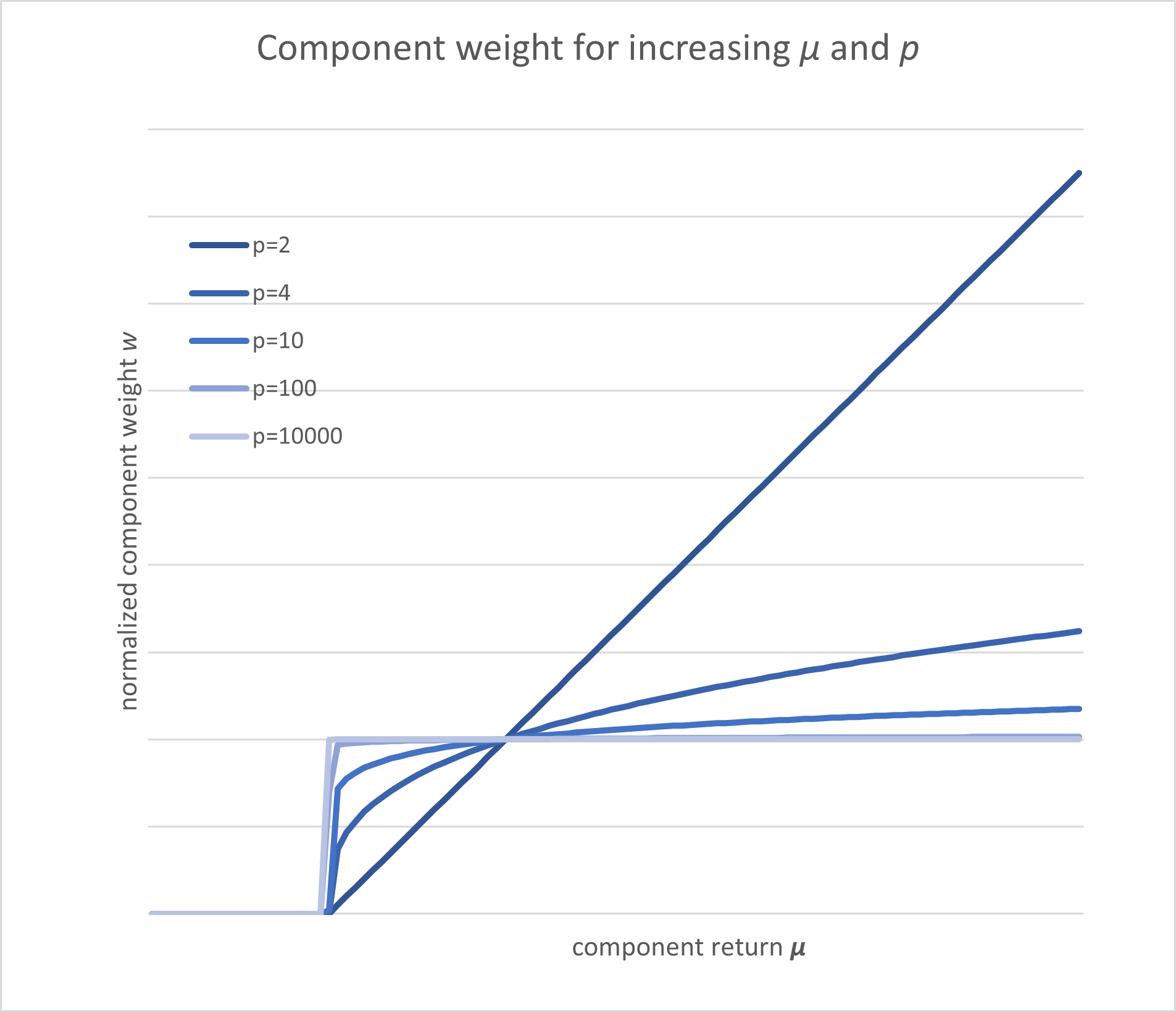}
\caption{Schematic representation of the portfolio weight $w^{(i)}$ as a function of its return $\mu^{(i)}$, for non-zero hurdle rate $r$ various values of $p$. }
\label{fig:weights}
\end{figure}

A graphical representation of the dependence of the weight of a component on its return is shown in Figure \ref{fig:weights}. 

As seen from Figure \ref{fig:weights}, the weight/return diagram has a simple limiting behaviour for large values of $p$; it approximates a step function with the step set at the hurdle rate $r$.

This is immediately obvious by examining the functional form of equation (\ref{ICAweights}); as $p$ tends to inifinity, the exponent $1/(p-1)$ tends to zero and $\mu^{1/(p-1)}$ tends to 1. 

While this observation is mathematically trivial, it has deep implications for portfolio management.

There is a well known trading heuristic, as cited by Giller \cite{giller}:
\begin{displayquote}
\emph{Don’t trade if the signal is too small. If it is large enough, buy a fixed size.}
\end{displayquote}

The reason is that a portfolio composed of equally weighted independent components is a perfectly diversified portfolio. It has a variance that decreases as $1/N$, and excess kurtosis that decreases as either $1/N$ or $1/N^{2}$, depending on whether the components are only orthogonal, or independent to higher orders \cite{rosenzweig}.

The limit of our equation (\ref{ICAweights}) as $p \rightarrow \infty$ directly recovers the step function implied by the heuristic. Noting that ICs are by construction normalized to the same volatility \cite{ hyva}, we can formulate it in words as:
\begin{displayquote}
\emph{Don’t trade if the expected return is smaller than the hurdle rate. If it is larger, buy a fixed volatility.}
\end{displayquote}

If hurdle rate $r=0$, the power law portfolio for $p\rightarrow \infty$ is  a perfectly diversified portfolio. Otherwise, the power-law portfolio will not  include the components that return less than the hurdle rate; it will therefore not have full $N$ components, and its variance decays as $1/N'$ for some smaller number of components $N' \le N$.

The portfolios which still retain non-trivial dependence on the return are generally not perfectly diversified, due to the fact that their components are not equally weighted in volatility. This is the case for all finite values of $p$.

On the other hand, power-law portfolios come closer to being perfectly diversified  as $p$ increases. We thus get a useful further rule-of-thumb for interpreting the order $p$. For small $p$, the portfolios are very dependent on their returns, at the expense of diversification. As $p$ increases, the portfolios give up return in exchange for the benefit of diversification. In the limit of $p\rightarrow \infty$, the portfolios become perfectly diversified, and the return is only used to determine whether the holding is long, short or zero.

\section{What if they are Gaussian?}

While our primary motivation is to study portfolios with non-normal returns, it is still worthwhile to examine the case where each $IC^{(i)}$ is Gaussian with the mean $\mu^{(i)}$ and standard deviation $\sigma^{(i)}$. In that case, dropping the superscripts for a moment, we have a simple formula for the absolute central moments,
\begin{equation}
M_{p} = \frac{1 }{\sqrt{\pi}}\ 2^{p/2}\ \Gamma \left( \frac{p+1}{2} \right)  \sigma^{p} \label{GaussianMoment}
\end{equation}
where $\Gamma()$ denotes the Gamma function and $\sigma$ is the volatility of the component.

Incorporating (\ref{GaussianMoment}) into (\ref{ICAweights}) and returning the superscripts, we get
\begin{equation}
w^{(i)} \propto \frac{1}{\sigma^{(i)}} \left( \frac{\mu^{(i)}-r^{(i)}}{\sigma^{(i)}}\right)^{1/(p-1)} 2^{-p/2(p-1)}  \Gamma \left( \frac{p+1}{2} \right)^{-1/(p-1)}.
\label{GaussianWeights}
\end{equation}

In other words, each weight is, after its corresponding component is normalized to unit volatility (term $1/\sigma^{(i)}$), proportional to the Sharpe ratio of the component, raised to the power of $1/(p-1)$.

The constant term involving the Gamma function is just a proportionality constant which will in practice be over-riden by normalization. We can, however, still simplify it further to better understand the infinite order limit from the previous section.

Using the Striling's formula
$$
\Gamma(z) = \sqrt{\frac{2\pi}{ z}} \left( \frac{z}{e}  \right)^{z} \left( 1 + O\left( \frac{1}{z} \right) \right),
$$
we get
\begin{equation}
w^{(i)} \propto \frac{1}{\sigma^{(i)}} \left( \frac{\mu^{(i)}-r^{(i)}}{\sigma^{(i)}} \right)^{1/(p-1)} \frac{ e^{p/2(p-1)}} {p^{p/2(p-1)} 2^{(p+1)/2(p-1)} }
\label{StirlingWeights}
\end{equation}When $p$ is sufficiently large, $p/(p-1) \approx 1$ and  this further simplifies to
\begin{equation}
w^{(i)} \propto \frac{1}{\sigma^{(i)}} \left( \frac{\mu^{(i)}-r^{(i)}}{\sigma^{(i)}} \right)^{1/(p-1)} \sqrt{ \frac{ e} {2 } } \frac{1}{\sqrt{p}}.
\label{StirlingWeightsLimit}
\end{equation}

The weight includes normalization to unit volatility (term $1/\sigma^{(i)}$), and then allocation proportional to the Sharpe ratio to the power of $1/(p-1)$. The amplitude of the weight decays as $1/\sqrt{p}$, which is in practice over-riden by the  normalization of the weights.

Again, we can express the allocation as a heuristic in words for any given $p$:
\begin{displayquote}
\emph{Don’t trade if the expected return is smaller than the hurdle rate. If it is larger, normalize to unit volatility and buy a size in proportion to the Sharpe ratio to the power of $1/(p-1)$.}
\end{displayquote}
In the limit of $p$ going to infinity, this remains as in the previous section:
\begin{displayquote}
\emph{Don’t trade if the expected return is smaller than the hurdle rate. If it is larger, buy a fixed volatility.}
\end{displayquote}

\section{S\&P 500 stocks}

We looked at the same data set as in \cite{rosenzweig}, namely S\&P 500 stocks over a period of 12 years, from the 1st January 2007 until the 31st December 2018. To counteract the effects of stocks drifting in and out of the index over such a long time frame, we have divided the time frame into four buckets, each lasting three calendar years; from 1st January 2007 until 31st December 2009, from 1st January 2010 until 31st December 2012, from 1st January 2013 until 31st December 2015 and from 1st January 2016 until 31st December 2018. The basket for each bucket was selected as consisting of the index constituents on the last business day prior to the start of the bucket, and these stocks were followed until the end of the bucket. Any stock that was de-listed before the end of a bucket in which it appeared was deemed to have returned $0\%$ from its last trading day until the end of the bucket. There were no adjustments for stocks entering or leaving the index over the duration of any of the buckets.

\begin{figure}[p]
\centering
\includegraphics[width=7cm]{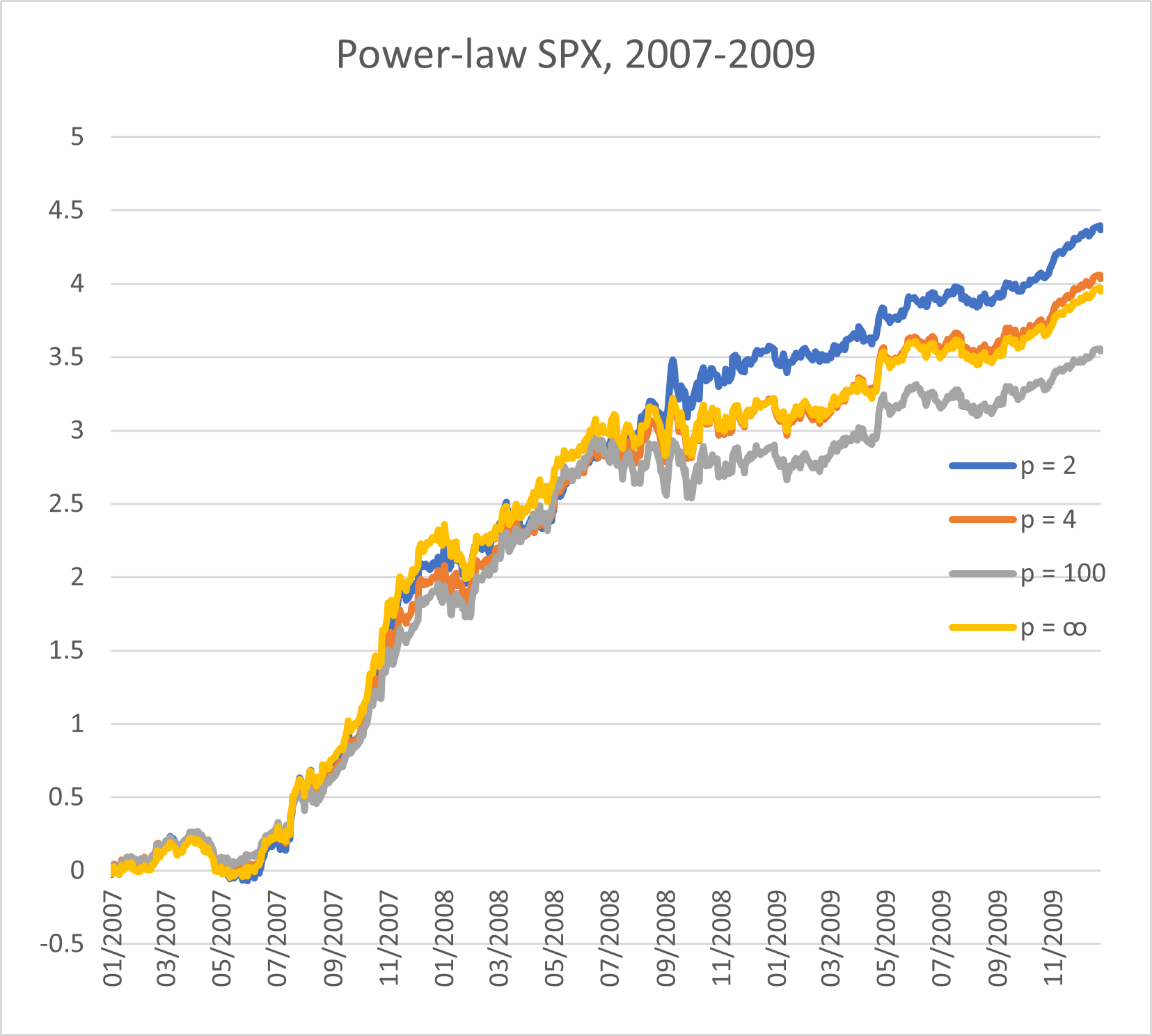}
\centering
\includegraphics[width=7cm]{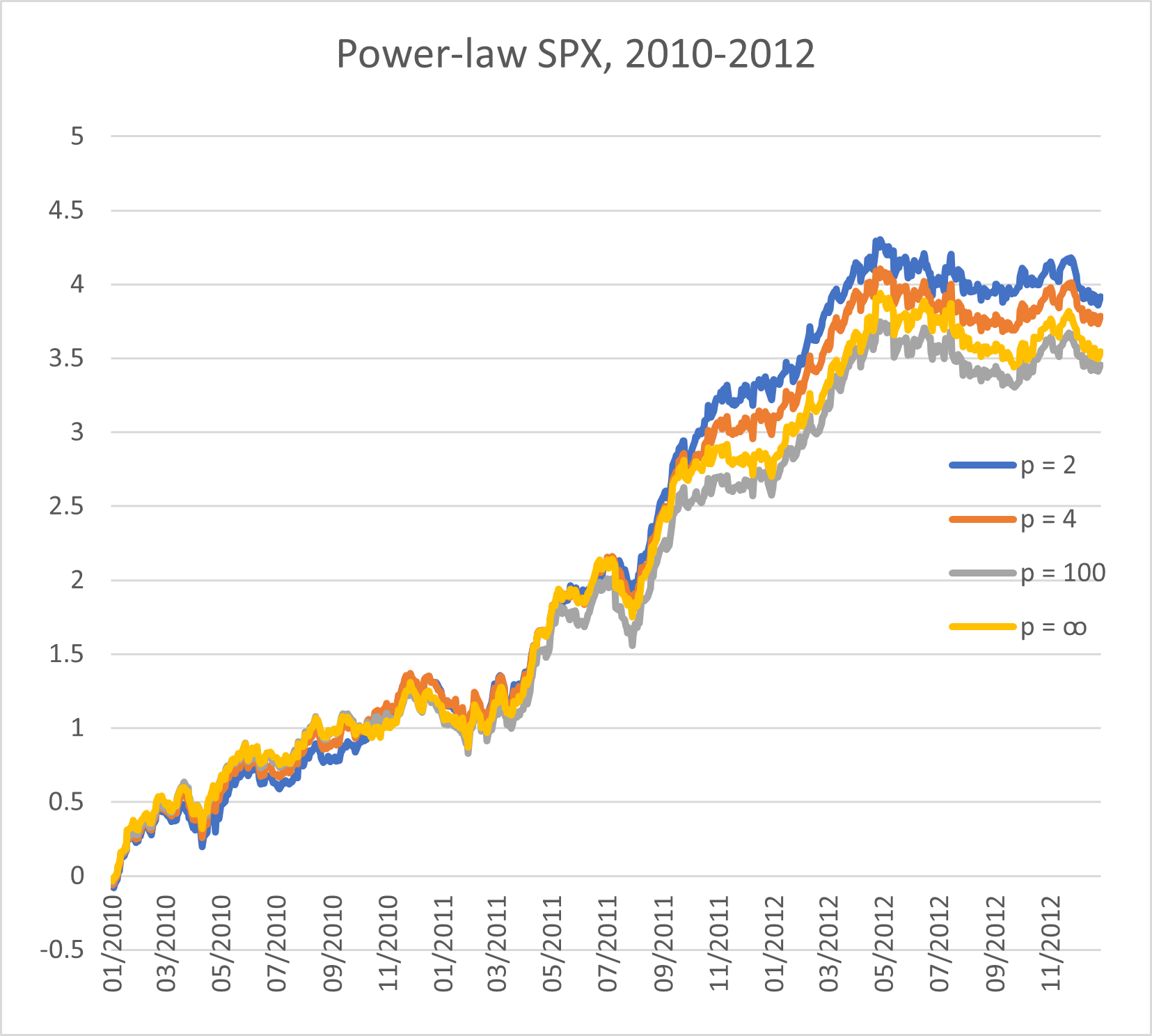}
\centering
\includegraphics[width=7cm]{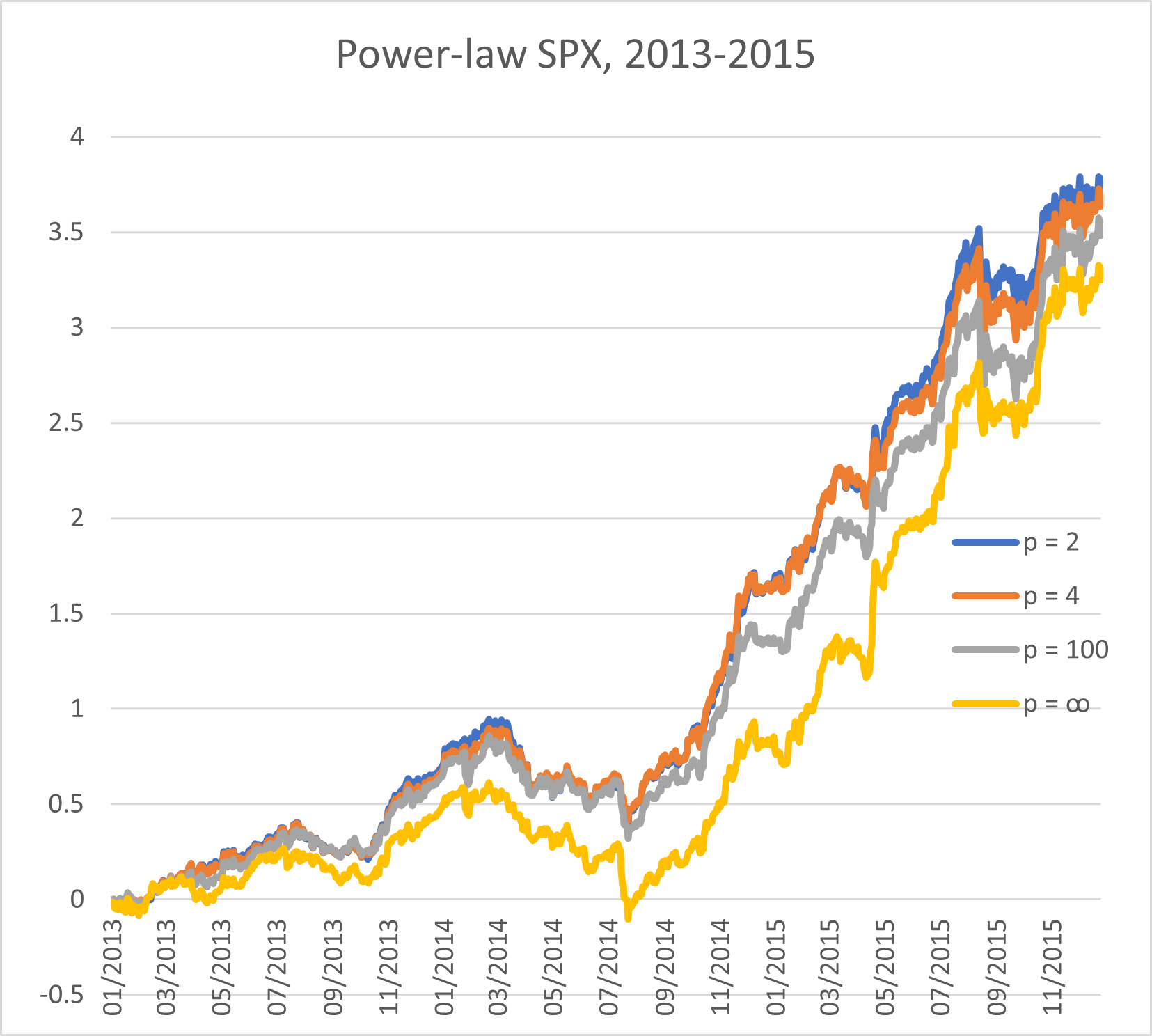}
\centering
\includegraphics[width=7cm]{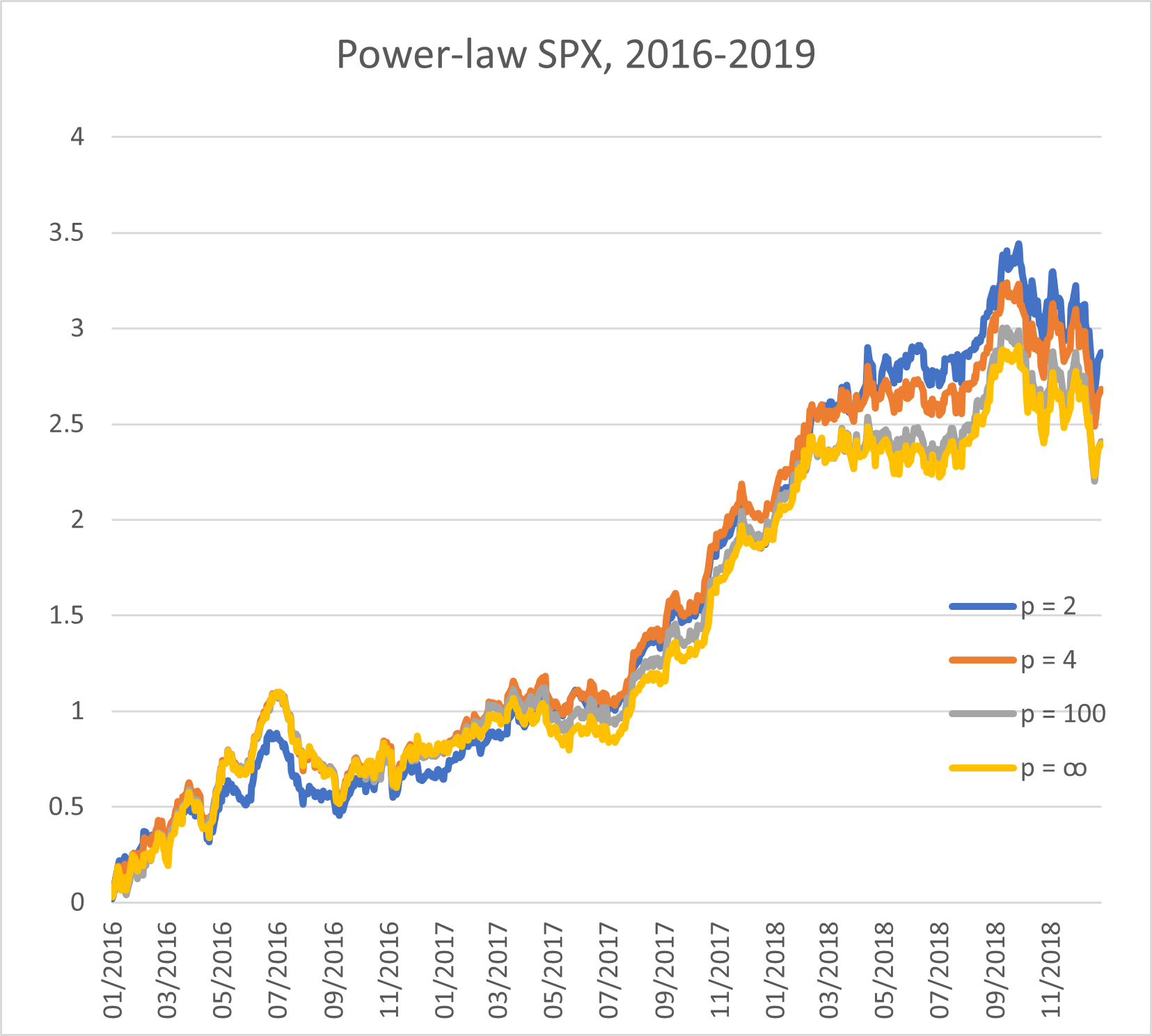}
\caption{Power-law portfolios for the S\&P 500 components for the 2007-2009, 2010-2012, 2013-2015 and 2016-2018 buckets with $p=2,4,100$ and $\infty$. The hurdle rate $r$ is set to 0,  portfolio weights are normalised to $\sum_{i} \left( w^{(i)} \right)^{2} = 1$, and \% returns are \% of  $\sqrt{\sum_{i} \left( w^{(i)} \right)^{2}}$. }
\label{fig:portfolio}
\end{figure}

We have extracted the first ten ICs and constructed the resulting power-law portfolios corresponding to $p=2,4,100$ and $\infty$. The performance of the resulting portfolios is shown in Figure \ref{fig:portfolio}.

As expected from the theoretical analysis, the Kelly portfolio for $p=2$ is the most aggressive in each bucket, having the highest weighting by return. The portfolios become incresingly less aggressive for increasing $p$. Perhaps counter-intuitively, $p=\infty$ is not always the least aggressive portfolio. In the two earliest buckets, 2007-2009 and 2010-2012, the portfolio for $p=100$ is less aggressive than the portfolio for $p=\infty$. 

This is not as surprising as it seems. The buckets 2007-2009 and 2010-2012 include the global financial crisis when stock returns were highly erratic, which was reflected in the high order return moments. The $p=\infty$ case is agnostic of  return moments, and it relies purely on diversification.

The portfolio statistics are shown in Table \ref{tab:stats}. The immediately obvious feature is that, in each bucket, the $p=2$ portfolio has the highest Sharpe Ratio, and the $p=4$ portfolio has the highest Fat-tailed Ratio. This is entirely unsurprising in light of the theoretical results above, since the $p=2$ portfolio by construction maximizes the Sharpe ratio, and the $p=4$ portfolio by construction maximizes the Fat-tailed Ratio. We did not show the $p=100$ power-law ratio, but, by construction, it is maximized by the $p=100$ portfolio. And for any other choice of $p$, the specific $p$-portfolio maximizes the $p$th power-law ratio. Note that there is no simple $p=\infty$ ratio to compare.

Looking at the correlations in Table \ref{tab:corr}, it is noticeable that all portfolios reproduce the same factors. Correlations are positive and high across the board. We see correlations occasionally dipping towards 80\% in a handful of places, always between $p=2$ and one of the higher order portfolios, either $p=100$ or $p=\infty$. Otherwise, they are comfortably above 90\%, and often above 95\%.

The differences in the Sharpe ratio between different portfolios in the same bucket are between 10 and 20\%, with 20\% being reached between the $p=2$ and either $p=100$ (in 2007-2009) or $p=\infty$ (in 2016-2018). Those particular buckets also seem to have the highest difference in the Fat-tailed ratio, this time in favour of the higher order portfolios. 

Differences in the Sharpe ratio of 10\% or less can be attributed to the perfect hindsignt that was used in our portfolio construction, and it is unlikely that they would translate into forward-looking portfolio construction in the real world.

The conclusion seems to be that, in low volatility environments, there is not much to choose between the portfolios for different values of $p$. In high volatility environments, however, we have a clear choice of whether to push the risk out of the volatility and into the tails (for $p=2$), or out of the tails and into the volatility (for large values of $p$). There appears to be no choice of $p$ that would predictably and simultaneously reduce all risk measures at the same time.

\section{Conclusions}

The method describeed here is a straightforward generalization of the Kelly criterion to non-Gaussian portfolios, obtained by moving the risk penalty from the second moment, variance, to an arbitrary $p$th absolute moment of the returns, for some $p \ge 2$.

By doing so, we  can significantly reduce the dependence of the portfolio weight of a component on its return. The resulting weight scales with return to the power of $1/(p-1)$, which is sub-linear when $p>2$. The resulting portfolio is better diversified than a corresponding Kelly portfolio, and less susceptible to the winner-takes-all problem  in which a handful of strongly performing components attract a lion's share of the capital. 

The diversification effect becomes stronger as $p$ increases. In the limit of $p$ going to infinity,  the weight becomes a simple 0-1 digital step function, whereby a component is assigned either a fixed weight, if its return exceeds a hurdle, or zero otherwise. This is the well known trading heuristic,
\begin{displayquote}
\emph{Don’t trade if the signal is too small. If it is large enough, buy a fixed size.}
\end{displayquote}
 By formalizing it, we have strengthened it to 
\begin{displayquote}
\emph{Don’t trade if the signal is smaller than the hurdle rate. If it is larger, buy a fixed volatility.}
\end{displayquote}

The resulting portfolios capture the same factors regardless of the chosen value of $p$. The choice of $p$ only affects the ultimate risk profile of the resulting portfolio.

There is no free lunch in finance, and this portfolio construction method is not a free lunch. By penalizing for moments of any given order $p$, we succeed in pushing the risk away from the $p$th moment, but it only moves into other moments. The portfolio construction method we present follows a simple logic:
\begin{itemize}
\item \emph{If you can hedge it, hedge it.}
\item \emph{If you can't hedge it, diversify it.}
\item \emph{If you can neither hedge nor diversify it, push it somewhere else.}
\end{itemize}

The hedging arises through the use of the Independent Component Analysis, which generates components within which individual assets hedge each other as far as possible. The diversification arises through weights given to the components, which generate as diversified a portfolio of independent components as specified for the given value of $p$.

The final step, of pushing risk away from the monitored moment into other moments is the most problematic.

On the more positive side, choosing a high value of $p$ results in risk being pushed out of the tails and into volatility, where it is the easiest to monitor. This is, arguably, preferable to Kelly portfolios which push risk out of volatility and into the tails, where it is more difficult to monitor.

A further benefit of the Independent Component Analysis used here as opposed to Principal Component Analysis is reduced dependence on pairwise asset correlations, which is recognised as a primary weakness in Kelly portfolios. By choosing components which are independent to all orders, as opposed to just orthogonal, we arguably recover more stable components which are less likely to arise due to sampling bias \cite{rosenzweig}. 

In summary, sub-linear power law portfolios constitute a powerful portfolio construction method which addresses some well known deficiencies of Kelly portfolios. In the form of digital, fixed-size-or-nothing allocation of the infinite order limit, it has already been a mainstay of real-world portfolio construction for decades, if not centuries. We here provide a simple rationale for its use, and we put it into a rational, objective framework.

\section*{Acknowledgments}

The author reports no conflicts of interest. The author alone is responsible for the content and writing of the paper.

\section*{Appendix - Power Law Formula}

This appendix outlines the proof of the power-law formula (\ref{ICAweightsnohurdle}). 

First, we focus on the cases where $p$ is an even integer, $p=2k$. Then, the absolute moment is the same as the signed moment, and we can drop the absolute value and expand the moment calculation. Using the multinomial theorem and independence of ICs, we get

\begin{equation}
\begin{array}{rl}
E \left| d\Pi \right| ^{p}  &= E \left(  d\Pi \right)^{2k} =  E \left(  \sum_{i=0}^{N} w^{(i)} d IC^{(i)}  \right)^{2k} =\\
&= E \left[  \sum_{k_{1}+k_{2}+...+k_{N}=2k} \prod_{i=1}^{N}   \left( w^{(i)} d IC^{(i)}  \right)^{k_{i}} \right]=\\
&=  \sum_{k_{1}+k_{2}+...+k_{N}=2k} \prod_{i=1}^{N}   \left( w^{(i)}  \right)^{k_{i}}   m^{(i)}_{k_{i}} 
\end{array}
\label{binomial}
\end{equation}
It is a standard result that powers of independent random variables are independent, and therefore all cross moments of independent random variables are separable \cite{hyva}.

In particular, we can also note that $m_{1}^{(i)}=0$ for all $i$ due to centrallity.

Then, substituting (\ref{binomial}) into (\ref{optimization}) and taking the first derivatives wrt each $w^{(i)}$, we get a system of equations of the form
\begin{equation}
\left(\mu^{(i)} - r^{(i)} \right)- P_{i}\left( w^{(i)} \right) = 0
\label{polyoptimization}
\end{equation}
where each $P_{i}$ has the following properties:
\begin{itemize}
\item  $P_{i}$ is a polynomial of order $2k-1$
\item $P_{i}(0)=0$
\item the leading coefficient of each $P_{i}$ is $m_{2k}^{(i)}$, which is non-negative, and independent of $w^{(j)}$ for all $j\ne i$.
\end{itemize}

If $m_{2k}^{(i)}=0$, the weight of that component can be increased without incurring the penalty; the maximum of (\ref{optimization}) is then reached at the boundary,  $w^{(i)}=1,$  $w^{(j)}=0$ for  $j\ne i$.

In general, if all $m_{2k}^{(i)}>0$,   (\ref{polyoptimization}) can have up to $2k-1$ real roots for each $i$, corresponding to up to $(2k-1)^{N}$ maxima and minima for (\ref{optimization}). We specifically focus on the largest zero of (\ref{polyoptimization}) for each $i$, which we denote $\hat{w}^{(i)}$.

Then, if $\mu^{(i)} - r^{(i)} > 0$, the following propositions hold:
\begin{enumerate}
\item  $\hat{w}^{(i)} \ge 0$ for all $i$;
\item ${\hat{\bf w}} = (\hat{w}^{(i)})$ is a local maximum of (\ref{optimization});
\item if all coefficients of each $P_{i}$ are non-negative, then ${\hat{\bf w}}$ is the only non-negative maximum of (\ref{optimization});
\item for each $i$, there is a lower bound $b^{(i)} < 0$ such that, if all coefficients of  each $P_{i}$ are greater than $b^{(i)}$, then  ${\hat{\bf w}}$ is the global non-negative maximum of (\ref{optimization}).
\end{enumerate}

Briefly, 1 and 2 are direct consequences of the Descartes' rule of signs; 3 follows from the fact that non-negativity of coefficients makes each $P_{i}$ a monotonically increasing function on $[0,\infty)$; and 4 follows from $3$ by continuity.

Note that 3 (and therefore 4) is trivially satisfied if the ICs are all normally distributed, or if they folllow  any symmetric distribution with finite moments up to order $2k$. 

While one can not generally guarantee that any return distribution of interest in finance will always satisfy 4, most of them do.  We can therefore quite generally accept ${\hat{\bf w}} $ as a universal local maximum, and a nearly-universal global maximum of (\ref{optimization}). 

We can finally come back to the assumption that $\mu^{(i)} - r^{(i)} > 0$ above. ICs are generally sign-agnostic; ICA decomposition is unaffected by the transformation $IC^{(i)} \mapsto -IC^{(i)}$; in finance terms, ICs are generally long-short portfolios with no obvious sign. However, given the difference in the funding cost of long and short positions, the funding rate of $-IC^{(i)}$ is generally not  $-r^{(i)}$.

Therefore, if the assumption that  $\mu^{(i)} - r^{(i)} > 0$ can not be satisfied by changing the sign of $IC^{(i)}$, i.e. if  $IC^{(i)}$ can not cover its cost of funding on either the long or the short side, then it can not contribute positive return to the portfolio; therefore its best possible contribution to the maximum of (\ref{optimization}) is achieved by setting its weight to zero. Mathematically, if the constant term in (\ref{polyoptimization}) is negative, Descartes' rule of signs no longer implies that  (\ref{polyoptimization}) has a positive zero, and therefore the maximum is reached on the boundary, $\hat{w}^{(i)} = 0$.

It now remains to estimate $\hat{w}^{(i)}$, and thereby ${\hat{\bf w}}$, for  components that satisfy $\mu^{(i)} - r^{(i)} > 0$. This is straightforward;  $\hat{w}^{(i)}$ has a known Puiseux series expansion \cite{algebraic}
\begin{equation}
w^{(i)} = \left( \frac{\mu^{(i)} - r^{(i)}}{m_{2k}^{(i)}} \right)^{1/(2k-1)} + C_{0}^{(i)} +  C_{1}^{(i)}\left( \mu^{(i)} - r^{(i)} \right)^{-1/(2k-1)} +...  \label{minhelper}.
\end{equation}
Then, noting that $m_{2k}^{(i)} = M_{2k}^{(i)}$ and that $2k=p$, we get

\begin{equation}
w^{(i)} = \left( \frac{\mu^{(i)} - r^{(i)}}{M_{p}^{(i)}} \right)^{1/(p-1)} + C_{0}^{(i)} +  C_{1}^{(i)}\left( \mu^{(i)} - r^{(i)} \right)^{-1/(p-1)} +..., \label{minhelper1}
\end{equation}
, or
\begin{equation}
w^{(i)} = \left( \frac{\mu^{(i)} - r^{(i)}}{|m_{p}^{(i)}|} \right)^{1/(p-1)} + C_{0}^{(i)} +  C_{1}^{(i)}\left( \mu^{(i)} - r^{(i)} \right)^{-1/(p-1)} +..., \label{minhelper11}
\end{equation}

When $p$ is not an even integer, the situation is somewhat more complicated. We can verify that there is a local extremum of (\ref{optimization}) satisfying (\ref{minhelper1}) by substituting (\ref{minhelper1}) into (\ref{optimization}) and setting the first derivatives wrt each $w^{(i)} $ to zero.

However, proving global properties of (\ref{minhelper1}) is  more difficult. We generally do it by approximating the moment function with a polynomial of even order, and then using the polynomial analysis above,

One way forward relies on  noting that, if $p$ is sufficiently large, we can always write $p=2k(1+\epsilon)$ for some integer $k$ and $\epsilon =  (p-2k)/2k  \ll 1$; we can then expand (\ref{optimization}) in powers of $\epsilon$ and revert to the polynomial analysis above,  resulting in  (\ref{minhelper1})  plus an error term of the order $O(\epsilon)=O(1/p)$.

This leaves us with the case when when $p$ is not an even integer and not large. Whether or not the local maximum (\ref{minhelper1}) is generally the global maximum of (\ref{optimization})  in this case, and under what conditions, is currently, to the best knowledge of the author, unknown.

Formulation (\ref{optimization1}) has some simple solutions for non-even values of $p$. For $p=3$, all cross terms in the polynomial are of the form $m^{(j)}_{1}m^{(k)}_{2}$ and they are all equal to zero, given the vanishing first central moments $m^{(j)}_{1}$. The formula (\ref{minhelper11}) is therefore trivially satisfied at the leading order, with $C_{0}=C_{1}=....=0$.

Higher non-even values of $p$ are not as simple. For $p=5$, for example, we get  the cross term  $m^{(j)}_{2}m^{(k)}_{3}$ which might be zero, positive or negative. This, after taking the absolute value, thus increases the number of potential maxima, making the construction of  simple analytical solutions incresasingly difficult. Once $p$ becomes sufficiently high, however, the asymptotics  $p=2k(1+\epsilon)$ starts to work.

\begin{table}[p]
\begin{tabular}{ c |l | r | r | r | r |}
Bucket & Statistic, annualized & $p=2$ & $p=4$ & $p=100$ & $p=\infty$ \\ \hline
2007-2009 &Return &	146.16\% &	135.07\% &	118.47\% &	132.24\% \\
& Standard Deviation &	57.85\% &	57.85\% &	57.85\% &	57.85\% \\
& Kurtosis &	2.537 &	1.611 &	1.488 &	1.407 \\
& Sharpe Ratio &	2.527 &	2.335 &	2.048	& 2.286 \\
& Fat-tailed Ratio &	0.832 &	0.943	& 0.927 &	0.979 \\ \hline
2010-2012 & Return &	130.86\% &	126.44\% &	115.53\% &	118.50\% \\
&Standard Deviation &	57.81\% &	57.81\% &	57.81\% &	57.81\% \\
& Kurtosis &	1.135 & 0.580 & 0.824 & 1.324  \\
& Sharpe Ratio &	2.263 &	2.187	& 1.998 &	2.050 \\
& Fat-tailed Ratio &	1.048	& 1.296 &	1.119	& 0.964\\ \hline
2013-2015 & Return &	123.95\% &	121.55\% &	116.44\% &	108.58\% \\
& Standard Deviation &	57.80\% &	57.81\% &	57.81\% &	57.81\% \\
& Kurtosis &	3.888 & 3.758 & 4.440 & 7.210 \\
& Sharpe Ratio &	2.145	& 2.103 &	2.014 &	1.878 \\
& Fat-tailed Ratio &	0.683 &	0.686 &	0.640 &	0.532 \\ \hline
2016-2018 & Return &	96.08\% &	89.68\% &	80.54\% &	80.17\% \\
& Standard Deviation &	57.81\% &	57.81\% &	57.81\% &	57.81\% \\
& Kurtosis &	2.425	& 1.365 &	1.625 &	1.439 \\
& Sharpe Ratio &	1.662 &	1.551 &	1.393 &	1.387 \\
& Fat-tailed Ratio &	0.734 &	0.869 &	0.791 &	0.823 \\ \hline
\end{tabular}
\caption{Return statistics for the 2007-2009, 2010-2012, 2013-2015 and 2016-2018 buckets with $p=2,4,100$ and $\infty$. The hurdle rate $r$ is set to 0, portfolio weights are normalised to $\sum_{i} \left( w^{(i)} \right)^{2} = 1$, and \% returns are \% of  $\sqrt{\sum_{i} \left( w^{(i)} \right)^{2}}$. }
\label{tab:stats}
\end{table}

\begin{table}[p]
\begin{tabular}{ c |l | r | r | r | r |}
Bucket &  & $p=2$ & $p=4$ & $p=100$ & $p=\infty$ \\ \hline
2007-2009 &$p=2$ &	100.00\% &	92.41\% &	81.06\% &	90.49\%\\
&  $p=4$ &	& 100.00\% & 	96.11\% &	98.00\% \\
& $p=100$ & & &	100.00\% &	95.88\% \\
&  $p=\infty$ &	  &	  &	 	& 100.00\%\\ \hline
2010-2012 &$p=2$ &	100.00\% &	96.62\% &	88.29\% &	90.55\% \\
&  $p=4$ &	& 100.00\% &  96.63\% &	95.26\% 	  \\
& $p=100$ & & &	100.00\% &	96.77\% \\
&  $p=\infty$ &	  &	  &	 	& 100.00\%\\ \hline
2013-2015 &$p=2$ &	100.00\% &	98.06\% &	93.94\% &	87.60\%\\
&  $p=4$ &	& 100.00\% & 	97.77\% &	89.17\%  \\
& $p=100$ & & &	100.00\% &	93.31\% \\
&  $p=\infty$ &	  &	  &	 	& 100.00\%\\ \hline
2016-2018 &$p=2$ &	100.00\% &	93.33\% &	83.82\% &	83.44\%\\
&  $p=4$ &	& 100.00\% & 	97.18\% &	95.91\%  \\
& $p=100$ & & &	100.00\% &	97.73\% \\
&  $p=\infty$ &	  &	  &	 	& 100.00\%\\ \hline
\end{tabular}
\caption{Return correlations for the 2007-2009, 2010-2012, 2013-2015 and 2016-2018 buckets with $p=2,4,100$ and $\infty$. The hurdle rate $r$ is set to 0.}
\label{tab:corr}
\end{table}


\begin{thebibliography}{1}

\bibitem{ahmadi} Ahmadi-Javad, A. and Fallah-Tafti, M. 2017. {Portfolio optimization with entropic Value-at-Risk} https://arxiv.org/ftp/arxiv/papers/1708/1708.05713.pdf
\bibitem{avellaneda} Avellaneda, M. 2019. {Hierarchical PCA and applications to portfolio management}, {https://ssrn.com/abstract=3467712 or http://dx.doi.org/10.2139/ssrn.3467712 }
\bibitem{algebraic} Basu, S., Pollack, R., Roy, M.-F. 2006. \emph{ Algorithms in Real Algebraic Geometry. Algorithms and Computations in Mathematics 10 (2nd ed.)}. Springer-Verlag. doi:10.1007/3-540-33099-2. ISBN 978-3-540-33098-1.
\bibitem{cajas} Cajas, D. 2021. {Entropic portfolio optimization: A disciplined convex programming Framework} {https://ssrn.com/abstract=3792520}
\bibitem{follmer} Föllmer, H.; Schied, A. 2002. {Convex measures of risk and trading constraints} \emph{Finance and Stochastics}. 6 (4): 429–447. doi:10.1007/s007800200072.
\bibitem{giller} Giller, G. 2008. {Frictionless asset allocation with elliptically symmetric distributions of returns}, {https://ssrn.com/abstract=1300671}
\bibitem{hyva} Hyv\"{a}rinen, A.  (2013) {Independent component analysis: recent advances}, \emph{Philosophical Transactions: Mathematical, Physical and Engineering Sciences}. 371
\bibitem{lintner} Lintner, J. 1965. {The valuation of risk assets and the selection of risky investments in stock portfolios and capital budgets},\emph{ The Review of Economics and Statistics}. 47 (1): 13–39. doi:10.2307/1924119. JSTOcR 1924119.
\bibitem{litterman} Litterman, R. \& Scheinkman, J. 1991. { Common  factors  affecting  bond  returns}, \emph{The Journal of Fixed Income}.
\bibitem{mark1} Markowitz, H.M. 1952. {Portfolio selection}, \emph{The Journal of Finance}. 7 (1): 77–91. doi:10.2307/2975974. JSTOR 2975974.
\bibitem{mark2}  Markowitz, H.M. 1956. {The optimization of a quadratic function subject to linear constraints}, \emph{Naval Research Logistics Quarterly}. 3 (1–2): 111–133. doi:10.1002/nav.3800030110.
\bibitem{rosenzweig} Rosenzweig, J. 2021. {Fat-tailed factors}, https://arxiv.org/abs/2011.13637
\bibitem{sharpe}  Sharpe, W.F 1964. {Capital asset prices: A theory of market equilibrium under conditions of risk}, \emph{Journal of Finance}. 19 (3): 425–442. doi:10.2307/2977928. hdl:10.1111/j.1540-6261.1964.tb02865.x. JSTOR 2977928.
\bibitem{shkolnik}  Shkolnik, A.D., Goldberg, L. \& Bohn, J.R.  2016. {   Identifying   broad   and narrow    financial    risk    factors    with    convex    optimization},  https://ssrn.com/abstract=2800237  or http://dx.doi.org/10.2139/ssrn.2800237
\bibitem{tanzohren} Tan, V.W.C. \& Zohren, S. 2020. {Large non-stationary noisy covariance matrices: A cross-validation approach} https://arxiv.org/abs/2012.05757
\bibitem{tobin} Tobin, J. 1958. {Liquidity preference as behavior towards risk}, \emph{The Review of Economic Studies}. 25 (2): 65–86. doi:10.2307/2296205. JSTOR 2296205.

\end{thebibliography}
\end{document}